\documentclass{ptephy_mod}
\usepackage{amsmath,amssymb,amsthm,graphicx,physics,url}

\def\({\left(}
\def\){\right)}
\def\e{\mathrm{e}}
\def\a{\varkappa}
\def\b{\kappa}
\newcommand\bs[1]{\boldsymbol{\mathit{#1}}}

\begin{document}

\title{Distribution of the ratio of the two smallest eigenphases of the Jacobi unitary ensemble}
\author{\name{\fname{Shinsuke} \surname{M. Nishigaki}}{\ast}} 
\address{\affil{~}{
Graduate School of Natural Science and Technology, Shimane University, Matsue 690-8504, Japan}
\email{nishigaki@riko.shimane-u.ac.jp}}

\begin{abstract}
We compute the joint distribution of the two smallest eigenvalues of the Jacobi unitary ensemble
and the distribution of their ratio. We use a framework for calculating J\'{a}nossy densities based on the 
Tracy-Widom system of ordinary differential equations. For the cases studied here, the leading finite-$N$ 
correction to the ratio distribution, relative to its universal Bessel-kernel limit, is $\mathcal{O}(N^{-4})$. 
This behaviour reflects a nontrivial cancellation of lower-order corrections present in the Jacobi kernel. 
Eigenphase ratios may thus reveal subtle finite-size effects in quantum-chaotic systems 
with particle-hole or chiral symmetry.
As an exploratory arithmetic application, we compare the ratio distribution for the two lowest zeros of 
quadratic Dirichlet $L$-functions with the hard-edge limit for $\mathrm{Sp}(N)$. 
The data indicate that the $N_{\rm eff}^{-1}$ term vanishes, 
leaving an $\mathcal{O}(N_{\rm eff}^{-2})$ correction.
\end{abstract}

\subjectindex{random matrices, Jacobi ensemble, circular ensemble, gap ratio, L-function}

\maketitle

\section{Introduction}
The Jacobi unitary ensemble (JUE) of random matrices
has played important roles in quantum physics
(the conductance of disordered conductors \cite{Beenakker:1997}),
statistics (multivariate analysis of variance \cite{Muirhead:1982}),
and information theory (MIMO signal processing \cite{Dar:2013}).
Geometrically, the eigenvalue distributions of unitary matrices 
drawn from the invariant measures on seven Cartan classes of Riemannian symmetric spaces
take the form of Jacobi ensembles, 
which therefore provide `canonical' finite-size models for the respective symmetry classes.
The JUE may be contrasted with its noncompact analogue, the Gaussian unitary ensemble (GUE): 
Wigner's semicircular global eigenvalue density for the GUE,
$\bar{\rho}_{\mathrm{GUE}}(x)=(2N/\pi)(1-x^2)^{1/2}$,
contrasts with the `inverse semicircular' density
$\bar{\rho}_{\mathrm{JUE}}(x)=(N/\pi)(1-x^2)^{-1/2}$ for the JUE.
The endpoints of the spectra at $|x|=1$ are called soft and hard edges, respectively,
reflecting the powers $\pm1/2$ in the eigenvalue densities.
These characteristics determine the local eigenvalue correlations near the respective spectral edges. 
The logarithmic derivatives of the corresponding Fredholm determinants, 
which give the extreme-eigenvalue distributions, 
satisfy $\sigma$-Painlev\'{e} equations: PIV and PVI at finite $N$, 
and PII and PIII$'$ in the large-$N$ limit \cite{Tracy:1994a,Tracy:1994b,Tracy:1994c,Forrester:2004}
(summarised in \cite{Forrester:2010}, Chap.\,8.2.5).
Much work has since been devoted to finite-$N$ corrections to 
the extreme-eigenvalue statistics of the JUE and its confluent limit, 
the Laguerre unitary ensemble (LUE)
\cite{Haine:1999,Edelman:2016,Perret:2016,Bornemann:2016,Forrester:2019,Lyu:2021}.

In view of the aforementioned importance of the JUE for spectral statistics
at the hard edge, and of the successes of the integrable-systems
approach to its smallest eigenvalue distribution,
it is natural to extend these analyses to the joint distribution of 
the two smallest eigenvalues.
For the LUE 
this task was addressed by
Forrester and Hughes \cite{Forrester:1994} for the integer index case
and subsequently treated
by Forrester and Witte \cite{Forrester:2007} for generic values of the index.
At the soft edge of unitary ensembles, Witte et al.~\cite{Witte:2013} 
derived the joint distribution of the two largest eigenvalues.
Perret and Schehr \cite{Perret:2014} 
gave an alternative expression for their spacing distribution in the GUE.
In a different direction,
Oganesyan and Huse \cite{Oganesyan:2007} recognised a key advantage of ratios of 
consecutive spacings over the spacings themselves: 
such ratios require no unfolding by a system-specific mean eigenvalue density. 
Although they introduced spacing-ratio distributions for bulk spectra, 
these distributions have since become widely used as diagnostics of quantum chaos, 
often through the Wigner-like surmise (which approximates the $N\to\infty$ result
by that for $N=3$) \cite{Atas:2013a,Atas:2013b}. 
Their relevance at the hard edge, however, has long been overlooked.
We fill this gap by providing the corresponding
`canonical' finite-$N$ hard-edge distribution, 
using the framework developed by Tracy and Widom (TW) \cite{Tracy:1994c}
and applied to J\'{a}nossy densities in earlier works
\cite{Nishigaki:2021,Nishigaki:2024,Nishigaki:2026}.

This article is organised as follows:
Section 2 reviews the JUE and its hard-edge scaling.
Section 3 treats the case of integer Jacobi index $b$, 
and Section 4 develops the TW system for generic indices.
Section 5 presents the joint and ratio distributions and their finite-$N$ corrections. 
Section 6 compares these results with the ratio of the two lowest zeros of quadratic
Dirichlet $L$-functions. 
Section 7 discusses possible physical applications and summarises our conclusions. 
Classical results on Jacobi polynomials are collected in the appendix.

\section{Jacobi unitary ensemble}
\subsection{Eigenvalue distribution}
The Jacobi unitary ensemble of random Hermitian matrices $H=(H_{jk})_{j,k=1}^N$
with indices $(a, b)$, both greater than $-1$,
is defined by the probability measure for the matrix elements:
\begin{align}
d\mu(H)=\frac{1}{Z_N^{(a,b)}}
\prod_{i=1}^N dH_{ii} \prod_{j>k}^N 
d^2 H_{jk}\,
\Theta(1-|H|)
\det(\mathbb{I}-H)^a\det(\mathbb{I}+H)^b .
\label{JPD_JUE}
\end{align}
The Heaviside function $\Theta(1-|H|)$ indicates that
the domain of the matrices $H$ is restricted to 
those whose eigenvalues $\{x_1,\ldots,x_N\}$
all lie in $[-1, 1]$, and
$Z_N^{(a,b)}$ is the normalisation constant ensuring $\int d\mu(H)=1$.
Eq.\,(\ref{JPD_JUE}) leads to the probability measure of the eigenvalues,
\begin{align}
&d\mu(x_1,\ldots,x_N)_{\mathrm{JUE}_N^{(a,b)}}=
\frac{1}{S_N^{(a,b)}}
\prod_{i=1}^N \( dx_{i} w(x_i)\) \,
\Delta_N(\{x\})^2,
\nonumber\\
&w(x)=
\Theta(1-|x|)
(1-x)^a (1+x)^b,~~
\Delta_N(\{x\})=
\prod_{j>k}^N (x_j-x_k).
\label{EVD_JUE}
\end{align}
Here $S_N^{(a,b)}$ stands for the Selberg integral
ensuring  normalisation $\int\!\cdots\!\int_{-1}^1 d\mu(\{x\})=1$,
\begin{align}
S_N^{(a,b)}=N! \prod_{n=0}^{N-1} \frac{h_n^{(a,b)}}{\bigl(c_n^{(a,b)}\bigr)^2},
\label{Selberg}
\end{align}
where $h_n^{(a,b)}$ and $c_n^{(a,b)}$ are defined in (\ref{norm}).
Upon a change of variables $x_i=\cos \theta_i$,
Eq.\,(\ref{EVD_JUE}) for several particular values of $(a,b)$ (see Table \ref{table1})
provides the joint distribution of
complex eigenvalues
$\{\e^{\pm i \theta_1},\ldots, \e^{\pm i\theta_N}\}$
of Haar-distributed unitary matrices $U$ representing the Cartan embedding of 
compact Riemannian symmetric spaces of classes
D, B, C, and AIII of rank $N$
(\cite{Forrester:2010}, Chap.\,3.7).
\begin{table}[h]
\centering
\begin{tabular}{|c|c|c|c|}
\hline
Class & Symmetric Space & $a$ & $b$ \\ \hline
D &$\mathrm{SO}(2N)$ & $-1/2$ & $-1/2$ \\ 
B &$\mathrm{SO}(2N+1)$ & $1/2$ & $-1/2$ \\ 
B &$\mathrm{O}^-(2N+1)$ & $-1/2$ & $1/2$ \\ 
D &$\mathrm{O}^-(2N+2)$ & $1/2$ & $1/2$ \\ 
C &$\mathrm{Sp}(N)$ & $1/2$ & $1/2$ \\ 
AIII &$\mathrm{U}(N+N')/\mathrm{U}(N)\times\mathrm{U}(N')$ & $N'-N$ & 0\\
\hline
\end{tabular}
\caption{Symmetric spaces and Jacobi unitary ensembles}
\label{table1}
\end{table}
In the context of level statistics, these random-matrix ensembles model
the time-evolution operators of quantum systems endowed
with one of the following involutive symmetries:
$\mathcal{C}U\mathcal{C}^{-1}=U$ for an anti-unitary $\mathcal{C}$
such that $\mathcal{C}^2=\mathbb{I}$ (D, B) or  $\mathcal{C}^2=-\mathbb{I}$ (C),
or $\mathcal{P}U\mathcal{P}^{-1}=U^\dagger$ for a unitary $\mathcal{P}$ (AIII),
while no anti-unitary $\mathcal{T}$ exists such that $\mathcal{T}U\mathcal{T}^{-1}=U^\dagger$.

\subsection{Jacobi kernel}
Following the standard orthogonal-polynomial treatment (see e.g.\,\cite{Forrester:2010}, Sect.\,5.1), 
the eigenvalues of $H$ form a determinantal point process (DPP).
The corresponding Jacobi kernel is given by 
\begin{align}
K(x,y)&=\sqrt{w(x)w(y)}\sum_{n=0}^{N-1} \frac{P_n^{(a,b)}(x)P_n^{(a,b)}(y)}{h^{(a,b)}_n}
\label{K_Jacobi_sum}\\
&=C_N^{(a,b)}
\sqrt{w(x)w(y)}
\frac{P_N^{(a,b)}(x) P_{N-1}^{(a,b)}(y)-P_{N-1}^{(a,b)}(x) P_N^{(a,b)}(y)}{x-y},
\label{K_Jacobi}
\\
C_N^{(a,b)}
&=\frac{c_{N-1}^{(a,b)}}{c_{N}^{(a,b)}h_{N-1}^{(a,b)}}
=\frac{N!\,\Gamma (N+a+b+1)}{2^{a+b}(2N+a+b) \Gamma(N+a) \Gamma(N+b)},
\nonumber\\
K(x,x)&=
C_N^{(a,b)} w(x) \(P_{N}^{(a,b)}{}'(x)P_{N-1}^{(a,b)}(x)-P_{N-1}^{(a,b)}{}'(x)P_N^{(a,b)}(x)\):=\rho(x),
\label{rho_Jacobi}
\end{align}
where the Jacobi polynomials $P_n^{(a,b)}(x)$
are listed in (\ref{norm}).
By definition (\ref{K_Jacobi_sum}),
$K(x,y)$ satisfies the projection property $\int_{-1}^1 dy\,K(x,y)K(y,z)=K(x,z)$
and the normalisation condition
$\int_{-1}^1 dx\,K(x,x)=N$.
The spectrum in the vicinity of the hard edges $x=\pm 1$ is unfolded
by substituting $x= \pm\cos(z/N)$.
This scaling measures eigenphase distances from $0$ or $\pi$ and 
normalises the mean spacing of the $z$ variables to $\pi$ for $z\gg1$.
Under this hard-edge scaling, the Jacobi kernel (\ref{K_Jacobi}) and density (\ref{rho_Jacobi})
converge as $N\to\infty$ to the universal Bessel kernel and density
(see (\ref{impMH}) below),
\begin{align}
\frac{1}{N}\sqrt{\sin\frac{z}{N}\sin\frac{w}{N}}
K\(-\cos\frac{z}{N},-\cos\frac{w}{N}\)
&=\sqrt{zw}
\frac{z J_{b+1}(z) J_{b}(w) -J_{b}(z)w J_{b+1}(w)}{z^2-w^2}
+\mathcal{O}\Bigl(\frac{1}{N}\Bigr),
\label{K_Bessel}\\
\frac{1}{N}\sin\frac{z}{N}\,
\rho\(-\cos\frac{z}{N}\)
&=
\frac{z}{2} \(J_b(z)^2 -J_{b+1}(z)J_{b-1}(z) \)
+\mathcal{O}\Bigl(\frac{1}{N}\Bigr).
\label{rho_Bessel}
\end{align}
For $x=\cos(z/N)$, the index $b$ is replaced by $a$.

In Sect.\,4 we employ the two-component notation for the Jacobi kernel (\ref{K_Jacobi}),
\begin{align}
K(x,y)&=\frac{^t\Phi(x) J \Phi(y)}{x-y},~~
J=
\left[
\begin{array}{cc}
0 & 1\\
-1 & 0
\end{array}
\right],
\label{integrable}\\
\Phi(x)&=
\left[
\begin{array}{c}
\varphi(x)\\
\psi(x)
\end{array}
\right]
=\sqrt{C^{(a,b)}_N w(x)}
\left[
\begin{array}{c}
P_N^{(a,b)}(x) \\
P_{N-1}^{(a,b)}(x) 
\end{array}
\right].
\label{K_JUE}
\end{align}
By the differentiation and recurrence formulae for the Jacobi polynomials,
the component functions of $\Phi(x)$
are related by the coupled linear differential equations (LDEs):
\begin{align}
&m(x) \frac{d}{dx}\Phi(x)=
\left[
\begin{array}{rr}
A(x) & B(x)\\
-C(x) & -A(x)
\end{array}
\right]
\Phi(x),
\nonumber\\
&m(x)= (2N+a+b)(1-x^2),\ 
A(x)=\frac{-a^2+b^2-(2N+a+b)^2 x}{2},
\label{LDE}\\
&B(x)={2 (N+a) (N+b) },\ 
C(x)={2N (N+a+b)}.
\nonumber
\end{align}
From Eqs.\,(\ref{integrable})--(\ref{LDE}),
TW derived a third-order nonlinear differential equation 
(\cite{Tracy:1994c}, Eq.\,(5.62))
satisfied by the log-derivative of the Fredholm determinant (gap probability)
of the Jacobi kernel over an interval $I=[-1, s]$,
$\sigma(s)=(1-s^2)\frac{d}{ds} \log\det({1}-\bs{K}_{I})$.
The first integral of TW's differential equation leads to
a particular case of the $\sigma$-Painlev\'{e} VI equation \cite{Forrester:2004}.
\vspace{3mm}

\noindent
{\it Remark 1.}~ For the exceptional case with $a+b=0$,
the Jacobi kernel (\ref{K_Jacobi}) and  density  (\ref{rho_Jacobi})
are expanded in powers of $1/N^2$ rather than $1/N$,
so that the leading corrections to the Bessel kernel (\ref{K_Bessel})
and density  (\ref{rho_Bessel}) are of $\mathcal{O}(1/N^2)$.
This property is evident from the coefficient functions $m, A, B, C$ in (\ref{LDE})
that turn out to have definite parity in $N$.

\section{Integer index case}
\subsection{Preliminary: Laguerre ensemble}
The probability measure of the eigenvalues of the LUE is specified by
\[
d\mu(x_1,\ldots,x_N)_{\mathrm{LUE}_N^{(b)}}=
\frac{1}{S_N^{(b)}}
\prod_{i=1}^N \(dx_{i} \,x_i^b \,\e^{-x_i} \,\Theta(x_i)\)
\Delta_N(\{x\})^2
~~(b>-1),
\]
where $S_N^{(b)}$ 
ensures the normalisation $\int\!\cdots\!\int_{0}^\infty d\mu(\{x\})=1$.
As the Jacobi polynomials reduce to the Laguerre polynomials in the confluent limit
at the 
endpoint of the Jacobi support, 
\[
P_n^{(a,b)}(-1+{2x}/{a})\stackrel{a\to\infty}{\longrightarrow} 
(-1)^n L_n^{(b)}(x),
\]
both ensembles lead to the same Bessel kernel in the hard-edge scaling limit.
Forrester and Hughes \cite{Forrester:1994} showed that when the index $b$ 
takes a non-negative integer value,
the joint distribution of the two smallest eigenvalues $(t, s)$ of LUE$^{(b)}_N$
is expressed as a $(b+2)\times (b+2)$ determinant.
For the simplest case with $b=0$, it reads
\begin{align}
P_{\mathrm{L}}(t,s)=
\e^{-t-(N-1)s}(t-s)^2  
\(L_{N-1}^{(2)}(t-s) L_{N-2}^{(2)\prime}(t-s)
-L_{N-2}^{(2)}(t-s) L_{N-1}^{(2)\prime}(t-s)\).
\label{P12_Laguerre0}
\end{align}
By using the Mehler-Heine asymptotic expansion
of Laguerre polynomials \cite{Frenzen:1988}
\[
\frac{n!}{\Gamma(n+b+1)}
L_n^{(b)}\!\left(\frac{x^2}{4n}\right)
=
\left(\frac{x}{2}\right)^{-b}
\left\{
J_b(x)-\frac{x^2}{8n}J_{b+2}(x) +\mathcal{O}\Bigl(\frac{1}{n^2}\Bigr)
\right\},
\]
they derived
the joint distribution of
the two smallest eigenvalues
in the hard-edge scaling limit
after setting $t=z^2/(4N)$ and $s=w^2/(4N)$.
Namely, for the simplest case with $b=0$, Eq.\,(\ref{P12_Laguerre0}) leads to
\begin{align}
P_{\mathrm{B}}(z,w)
&=
\lim_{N\to\infty}
\frac{z}{2N}\frac{w}{2N}
P_{\mathrm{L}}\left(\frac{z^2}{4N}, \frac{w^2}{4N}\right)
\nonumber\\
&=
\frac{z w}{4}
\e^{-w^2/4}
\( I_2(\sqrt{w^2-z^2})^2-I_1(\sqrt{w^2-z^2})I_3(\sqrt{w^2-z^2}) \).
\label{P12_Bessel0}
\end{align}
In the next subsection, we adapt their treatment of $\mathrm{LUE}_N^{(b)}$ 
to $\mathrm{JUE}_N^{(a,b)}$ with nonnegative integer $b$ 
and  derive the distribution of the ratio of the corresponding unfolded eigenphases.

\subsection{Jacobi ensemble}
We denote the two smallest eigenvalues of
JUE$^{(a,b)}_N$
by $t$ and $s$, respectively.
We obtain the joint distribution of $(t,s)$ by conditioning two of 
the $N$ eigenvalues to lie at $t$ and $s$ and restricting
the remaining integration variables in (\ref{EVD_JUE}) to $[s,1]$:
\begin{align}
P_{\mathrm{J}}(t,s)=&\frac{N(N-1)}{S_N^{(a,b)}}
 (1-t)^a (1-s)^a (1+t)^b (1+s)^b(s-t)^2
\nonumber\\
\times&\int\!\cdots\!\int_s^1 \prod_{i=1}^{N-2}
\(dx_i (1-x_i)^a (1+x_i)^b (t-x_i)^2 (s-x_i)^2\) 
\Delta_{N-2}(\{x\})^2.
\label{P12a}
\end{align}
The affine change of variables $x_i=\((1-s)y_i+1+s\)/2$ 
transforms the above $(N-2)$-fold integral back to the JUE form,
\begin{align}
\(\frac{1-s}{2}\)^{(N-2)(N+a+b+2)}&
\!\int\!\cdots\!\int_{-1}^1 \prod_{i=1}^{N-2}
\(dy_i(1-y_i)^a (1+y_i)^2 (y_i-u)^b (y_i-v)^2 \) 
\Delta_{N-2}(\{y\})^2,
\nonumber\\
&
u=-\frac{3+s}{1-s},~~
v=-\frac{1+s-2t}{1-s}.
\label{P12b}
\end{align}
Up to the normalisation factor $S^{(a,2)}_{N-2}$,
this integral is the expectation of the product of
characteristic polynomials
$\prod_{\alpha=0}^{b+1}\prod_{i=1}^{N-2}(u_\alpha-y_i)$
in JUE$^{(a, 2)}_{N-2}$,
in the partial confluent limit
$u_0,\ldots ,u_{b-1} \to u$ and $u_{b}, u_{b+1}\to v$.
The Christoffel formula states that
this expectation is expressed
in a determinantal form (see e.g. \cite{Szego:1939}, Sect.\,2.5).
In the present case, this gives
\begin{align}
\mathbb{E}
\Biggl[\prod_{\alpha=0}^{b+1}
\prod_{i=1}^{N-2}(u_\alpha-y_i)\Biggr]_{\mathrm{JUE}^{(a,2)}_{N-2}}
\!\!=
\frac{\det\left[P^{(a,2)}_{N-2+j}(u_k)\right]_{j,k=0}^{b+1}}{
\prod_{j=0}^{b+1}c_{N-2+j}^{(a,2)}\,
\Delta_{b+2}(\{u\})}.
\label{cn}
\end{align}
In this partial confluent limit, the determinant formula reduces to a Wronskian determinant:
\begin{align}
\mathbb{E}
\Biggl[\prod_{i=1}^{N-2}(u-y_i)^b(v-y_i)^2\Biggr]_{\mathrm{JUE}^{(a,2)}_{N-2}}
\!\!=
\frac{\det
\left[
\left[\partial_u^k P^{(a,2)}_{N-2+j}(u)\right]_{{j=0,\ldots,b+1 \atop k=0,\ldots,b-1 }}
\left[\partial_v^k P^{(a,2)}_{N-2+j}(v)\right]_{{j=0,\ldots,b+1 \atop k=0,1}}
\right]
}{\prod_{j=0}^{b+1}c_{N-2+j}^{(a,2)}\,
G(b+1) \,(u-v)^{2b}},
\label{Econfl}
\end{align}
where $G(b+1)=\prod_{k=0}^{b-1} k!$.
By combining Eqs.\,(\ref{P12a})--(\ref{Econfl})
with Eqs.\,(\ref{Selberg}) and (\ref{norm}),
one obtains
\begin{align}
P_{\mathrm{J}}(t,s)=&
(-1)^{b(N-2)}N(N-1)
\frac{2^{-N^2-(a+b)N+2a+4}}{
G(b+1) \prod_{j=0}^{b+1}c_{N-2+j}^{(a,2)}}
\frac{S_{N-2}^{(a,2)}} {S_{N}^{(a,b)}} 
\nonumber\\
\times& 
(1-t)^a (1-s)^{N^2+(a+b) N-a-4} (1+t)^{-b}(1+s)^b  (s-t)^2 
\nonumber\\
\times& 
\det
\left.
\left[
\left[\partial_u^k P^{(a,2)}_{N-2+j}(u)\right]_{{j=0,\ldots,b+1 \atop k=0,\ldots,b-1 }}
\left[\partial_v^k P^{(a,2)}_{N-2+j}(v)\right]_{{j=0,\ldots,b+1 \atop k=0,1}}
\right]
\right|_{u=-\frac{3+s}{1-s},~v=-\frac{1+s-2t}{1-s}}.
\label{P12_Jacobi}
\end{align}
In the simplest Legendre case $a=b=0$, corresponding to the Haar measure on
the complex Grassmannian manifold 
$\mathrm{U}(2N)/\mathrm{U}(N)\times\mathrm{U}(N)$,
Eq.\,(\ref{P12_Jacobi}) specialises to
\begin{align}
\frac{N(N^2-1)}{2^{N^2+1}}   (1-s)^{N^2-4}(s-t)^2
\( (N+2) P_{N-2}^{(0,2)}(v) P_{N-2}^{(1,3)}(v)-(N+1)P_{N-1}^{(0,2)}(v) P_{N-3}^{(1,3)}(v)  \).
\label{P12_Jacobi00}
\end{align}

The `unfolded eigenphase' variables $z$ and $w$ corresponding to $t$ and $s$
are defined by $t=-\cos (z/N)$ and $s=-\cos (w/N)$.
Accordingly, their joint distribution takes the form
\begin{align}
P_{\mathrm{J}}^{\mathrm{(unf)}}(z, w)=\frac{1}{N^2}\sin \frac{z}{N} \sin \frac{w}{N}
P_{\mathrm{J}}\(-\cos \frac{z}{N},-\cos \frac{w}{N}\).
\label{Pzw}
\end{align}
Using the improved Mehler-Heine expansion (\ref{impMH}), one finds that
the joint distribution 
obtained from Eqs.\,(\ref{P12_Jacobi}) and (\ref{Pzw}) converges, 
for generic $a$ and integer $b$, 
to the universal Bessel form as $N\to\infty$ (\cite{Forrester:2010}, Eq. (8.100)).
In particular, when $a=b=0$, 
Eq.\,(\ref{P12_Jacobi00}), after the transformation (\ref{Pzw}), 
converges to $P_{\mathrm B}(z,w)$ in Eq.\,(\ref{P12_Bessel0}).

Finally, the distribution $p_{\mathrm{J}}(r)$ of the ratio $r=z/w$
follows from $P_{\mathrm{J}}^{\mathrm{(unf)}}(z, w)$:
\begin{align}
p_{\mathrm{J}}(r)&=
\int_0^{N\pi}dw \int_0^{w}
dz\,
P_{\mathrm{J}}^{\mathrm{(unf)}}(z,w)\,
\delta(r-z/w)
=\int_0^{N\pi}
dw\, w\, P_{\mathrm{J}}^{\mathrm{(unf)}}(r w, w).
\label{Prr}
\end{align}

This analytic treatment has a limitation: it does not apply when $b$ is noninteger.
These cases include three important symmetry classes D, B, and C (see Table 1),
which are conjectured to govern the statistics of low-lying zeros in families of 
$L$-functions \cite{Katz:1999}, to be treated in Sect.\,6.
Instead of evaluating the Fredholm determinant numerically by quadrature 
\cite{Bornemann:2010a,Nishigaki:2015},
we address this limitation using the TW theory in the next section.

\section{Tracy-Widom theory}
\subsection{J\'{a}nossy density}
To make this article self-contained, 
we begin by reviewing how the TW system determines the Jánossy density,
following Ref.\,\cite{Nishigaki:2021}.

Suppose a kernel ${K}(x,y)$
satisfies the two properties $\int_{\mathcal{D}} dy\,{K}(x,y){K}(y,z)={K}(x,z)$
and $\int_{\mathcal{D}} dx\,{K}(x,x)=N$ on $\mathcal{D} \subseteq \mathbb{R}$.
Given that the $p$-point correlation function of a DPP 
governed by $K(x,y)$ is expressed as 
\begin{align}
{R}_p(x_1,\ldots,x_p)=\det[K(x_i, x_j)]_{i,j=1}^p,
\label{Rp}
\end{align}
conditioning one of the $N$ points to lie at $t$ and then removing it
gives the reduced Palm process. 
Its $(p-1)$-point correlation function
\[
\tilde{R}_{p-1|1}(x_1,\ldots,x_{p-1}|t)=\frac{{R}_{p}(x_1,\ldots,x_{p-1},t)}{R_1(t)}
\]
retains the determinantal form $\det\bigl[\tilde{K}(x_i,x_j)\bigr]_{i,j=1}^{p-1}$, 
with a transformed kernel given by
the Schur complement of the scalar block 
$K(t,t)~(t=x_p)$ of (\ref{Rp}) \cite{Shirai:2003},
\begin{align}
\tilde{K}(x,y)=K(x,y)-\frac{K(x,t)K(t,y)}{K(t,t)}.
\label{Ktransf}
\end{align}
It enjoys the requisite properties
$\int_{\mathcal{D}} dy\,\tilde{K}(x,y)\tilde{K}(y,z)=\tilde{K}(x,z)$,
$\int_{\mathcal{D}} dx\,\tilde{K}(x,x)=N-1$.
For $K(x,y)$ in the Christoffel-Darboux form (\ref{integrable}),
the relationship (\ref{Ktransf}) between the kernels is associated with a
rational $\mathrm{SL}(2)$ gauge transformation between the two-component functions,
\begin{align}
\tilde{\Phi}(x)=U(x)\Phi(x),\ \ 
U(x)^{\pm1}=
{1}\pm\frac{\Phi(t)\,^t\Phi(t)J}{K(t,t)(x-t)},
\ \ \det U(x)=1,
\label{Phitransf}
\end{align}
and the transformed kernel (\ref{Ktransf}) still assumes the Christoffel-Darboux form 
\[
\tilde{K}(x,y)=\frac{^t \tilde{\Phi}(x) J \tilde{\Phi}(y)}{x-y}.
\]

Suppose that the function ${\Phi}(x)$ satisfies
the coupled LDEs
\begin{align}
&m(x)\frac{d}{dx}
{\Phi}(x)
=
\left[
\begin{array}{rr}
A(x) & B(x)\\
-C(x) & -A(x)
\end{array}
\right]
{\Phi}(x)
~~\mbox{with polynomials}\ m,  A,  B, C.
\label{LDE_TW}
\end{align}
Then
the gauge-transformed function $\tilde{\Phi}(x)$
also satisfies the coupled LDEs of the same form,
\begin{align}
&\tilde{m}(x)\frac{d}{dx}
\tilde{\Phi}(x)
=
\left[
\begin{array}{rr}
\tilde{A}(x) & \tilde{B}(x)\\
-\tilde{C}(x) & -\tilde{A}(x)
\end{array}
\right]
\tilde{\Phi}(x)
~~\mbox{with polynomials}\ \tilde{m}, \tilde{A},  \tilde{B}, \tilde{C},
\label{LDE_N}
\end{align}
where the gauge-transformed coefficient functions are given by
\begin{align}
\frac{1}{\tilde{m}(x)}
\left[
\begin{array}{rr}
\tilde{A} (x)& \tilde{B}(x)\\
-\tilde{C} (x)& -\tilde{A}(x)
\end{array}
\right]
=
\frac{1}{m(x)}
U(x)
\left[
\begin{array}{rr}
{A} (x)& {B}(x)\\
-{C}(x) & -{A}(x)
\end{array}
\right]
U(x)^{-1}-U(x) \frac{d}{dx}U(x)^{-1}.
\label{gaugetransf}
\end{align}
In component notation,
the transformations (\ref{Phitransf}) and (\ref{gaugetransf}) take the following forms:
\begin{align}
&
\begin{array}{l}
{\displaystyle \tilde{\varphi}(x) =
 \varphi(x) - \frac{\b(\a \varphi(x) - \b \psi(x))}{x-t}}\\
{\displaystyle \tilde{\psi}(x)= 
\psi(x) - \frac{\a(\a \varphi(x) - \b \psi(x))}{x-t}}
\end{array}
~~\mbox{with}~~
\begin{array}{l}
{\displaystyle \b:=\frac{\varphi(t)}{\sqrt{K(t,t)}}}\\
{\displaystyle \a:=\frac{\psi(t)}{\sqrt{K(t,t)}}}
\end{array} ,
\label{tildephipsi}\\
&~\tilde{m}(x)=(x-t)^2{m}(x),
\nonumber\\
&~\tilde{A}(x)=
(x-t)^2{A}(x)
+(x-t)(\a^2 {B}(x)-\b^2 {C}(x))
-\b \a F(x),
\nonumber\\
&~\tilde{B}(x)=
(x-t)^2{B}(x)
-(x-t)2\b (\b {A}(x)+\a {B}(x))
+\b^2 F(x),
\label{mABC_N}\\
&~\tilde{C}(x)=
(x-t)^2{C}(x)
+(x-t)2 \a (\a {A}(x)+\b{C}(x))
+\a^2 F(x),
\nonumber\\
&~~~~~~\mbox{where}\ \ F(x):=2 \b \a {A}(x)+\a^2 {B}(x)+\b^2 {C}(x)-m(x).
\nonumber
\end{align}
Tracy and Widom \cite{Tracy:1994c} established
that the Fredholm determinant $\det({1}-\bs{K}_I)$ of the integration operator
$\bs{K}_I$, acting on $L^2(I)$ over a union of intervals 
$I\subset\mathcal{D}$
with the Christoffel-Darboux kernel $K(x,y)$ (\ref{integrable}) satisfying (\ref{LDE_TW}),
is determined by a system of partial differential equations (PDEs)
with respect to the endpoints of $I$.
Accordingly, for the Jacobi kernel that satisfies (\ref{LDE}), 
the Fredholm determinant
$\det \bigl({1}-\tilde{\bs{K}}_I \bigr)$
of the transformed Jacobi kernel $\tilde{K}(x,y)$ can likewise be determined
from the TW system of PDEs.
By the Gaudin-Mehta theorem,
the conditional probability $\tilde{J}_1(t;I)$ that the interval $I$
contains no eigenvalue other than the one 
conditioned to lie at $t\in I$
is equal to $\det \bigl({1}-\tilde{\bs{K}}_I \bigr)$.
The J\'{a}nossy measure ${J}_1(t;I)dt$, defined as
the {\it unconditional} probability that the interval $I$
contains no eigenvalue except for one located in $[t, t+dt]$, is given by
$\tilde{J}_1(t;I)dt$ multiplied by the eigenvalue density $\rho(t)=K(t,t).$
For the single-interval case with $I=[-1,s]$, 
the joint distribution of the smallest eigenvalue $t$ and the second smallest eigenvalue $s$
is given by
\begin{align}
{P}_{\mathrm{J}}(t,s)=
-\Theta(s-t)
\rho(t)
\frac{d}{d s} 
\det \bigl({1}-\tilde{\bs{K}}_{[-1,s]}\bigr).
\label{P12}
\end{align}

\subsection{Tracy-Widom system}
For a single interval $I=[-1,s]$, the TW system for the Fredholm determinant 
$\det \bigl({1}-\tilde{\bs{K}}_I \bigr)$ consists of a set of coupled nonlinear ODEs in $s$
involving the following quantities:
\begin{align}
R&
=(({1}-\tilde{\bs{K}}_I)^{-1} \tilde{K})(s,s)
\nonumber\\
&=
\tilde{K}(s, s)+
\int_I dx \, \tilde{K}(s, x)\tilde{K}(x, s)+
\iint_I dx dy\, \tilde{K}(s, x)\tilde{K}(x, y)\tilde{K}(y, s)+\cdots,
\nonumber\\
q_j&=(({1}-\tilde{\bs{K}}_I)^{-1}  x^j\tilde\varphi)(s)
\nonumber\\
&=
s^j \tilde{\varphi}(s)+
\int_I dx \, \tilde{K}(s, x) x^j \tilde{\varphi}(x)+
\iint_I dx dy\, \tilde{K}(s, x) \tilde{K}(x, y)y^j \tilde{\varphi}(y)+\cdots,
\label{Neumann}\\
p_j&=(({1}-\tilde{\bs{K}}_I)^{-1}  x^j\tilde\psi)(s),
\nonumber\\
u_j&=\int_I dx\, \tilde\varphi(x) (({1}-\tilde{\bs{K}}_{I})^{-1} x^j \tilde\varphi)(x),\ \ 
v_j=\int_I dx\, \tilde\varphi(x) (({1}-\tilde{\bs{K}}_{I})^{-1} x^j \tilde\psi)(x),
\nonumber\\
\tilde{v}_j&=\int_I dx\, \tilde\psi(x) (({1}-\tilde{\bs{K}}_{I})^{-1} x^j \tilde\varphi)(x),\ \ 
w_j=\int_I dx\, \tilde\psi(x)(({1}-\tilde{\bs{K}}_{I})^{-1} x^j  \tilde\psi)(x).
\nonumber
\end{align}
We regard $R, q_j, p_j, u_j, \ldots$ as functions of $s$ and 
use primes to denote differentiation with respect to $s$;
their parametric dependence on the fixed smallest eigenvalue $t$, 
through the transformed kernel $\tilde{K}(x,y)$ in Eq.\,(\ref{Ktransf}), is left implicit.
The nonuniversal TW coefficients $\mu_j, \alpha_j, \beta_j, \gamma_j$
are those of the polynomials in (\ref{LDE_N}), obtained by substituting (\ref{LDE}) into (\ref{mABC_N}),
\begin{align}
\tilde{m}(x)&=
 (2N+a+b)\(t^2-2 t x+(1-t^2) x^2+2 t x^3-x^4\)
~:=\sum_{j=0}^4 \mu_j x^j,
\label{mt}
\\
\tilde{A}(x)&=
\varkappa \kappa \(\varkappa \kappa (a^2-b^2)-2 \varkappa^2 (N+a) (N+b)-2 \kappa^2 N (N+a+b)+2N+a+b\)
\nonumber\\
&~~~
+\(2 \kappa^2 N (N+a+b)-2 \varkappa^2 (N+a) (N+b)\)t +\frac{b^2-a^2}{2} t^2
\nonumber\\
&+\Bigl(\varkappa^2 \kappa^2 (2N+a+b)^2+2 \varkappa^2 (N+a) (N+b)-2 \kappa^2 N (N+a+b)
\nonumber\\
&~~~
+ (a^2-b^2)t
-\frac{(2N+a+b)^2}{2}  t^2\Bigr)x
\nonumber\\
&+\Bigl(\frac{b^2-a^2}{2}-\varkappa \kappa (2N+a+b)+(2N+a+b)^2 t \Bigr)x^2
\nonumber\\
&
-\frac{ (2N+a+b)^2}{2} x^3
~:=\sum_{j=0}^3 \alpha_j x^j,
\label{At}
\end{align}
\begin{align}
\tilde{B}(x)&=
\kappa^2 \(\varkappa \kappa (b^2-a^2)+2 \varkappa^2 (N+a) (N+b)+2 \kappa^2 N (N+a+b)-(2N+a+b)\)
\nonumber\\
&~~~
+\kappa  \(\kappa (b^2-a^2)+4 \varkappa (N+a) (N+b)\)t+2  (N+a) (N+b)t^2
\nonumber\\
&+\(
\kappa^2 (a^2-b^2)
-\kappa^2(\varkappa \kappa+t) (2N+a+b)^2
-4 (\varkappa \kappa+t) (N+a) (N+b)
\)x
\nonumber\\
&+\(
\kappa^2 (2N+a+b)(2N+a+b+1)
+2 (N+a) (N+b)\)x^2
~:=\sum_{j=0}^2 \beta_j x^j,
\label{Bt}\\
\tilde{C}(x)&=
\varkappa^2 \(\varkappa \kappa (b^2-a^2)+2 \varkappa^2 (N+a) (N+b)+2 \kappa^2 N (N+a+b)-(2N+a+b)\)
\nonumber\\
&~~~
+\varkappa \(\varkappa   (a^2-b^2)-4 \kappa N (N+a+b)\)t +2 N (N+a+b) t^2
\nonumber\\
&+\(
-\varkappa^2 (a^2-b^2)
-\varkappa^2 (\varkappa\kappa-t) (2N+a+b)^2
+4 (\varkappa \kappa-t) N (N+a+b)
\)x
\nonumber\\
&+\(
-\varkappa^2 (2N+a+b-1) (2N+a+b)
+2 N (N+a+b)
\)x^2
~:=\sum_{j=0}^2 \gamma_j x^j,
\label{Ct}
\end{align}
as functions of $N, a, b$, and $t$, 
which also determine $\kappa$ and $\varkappa$ by (\ref{tildephipsi}).
With these preparations, 
the relevant equations of Ref.\,\cite{Tracy:1994c}, 
namely Eqs.\,(2.8), (2.12)--(2.16), (2.18), (2.25), (2.26), (2.31), and (2.33), 
yield the following TW system, which determines
$R(s)=p_0 q_0'-q_0 p_0'$:
\begin{align}
\tilde{m}(s)
q_0'
&=
\sum_{j=0}^3\(\alpha_j+\sum_{k=0}^2 \alpha_{j+k+1}v_k+\sum_{k=0}^1 \gamma_{j+k+1}u_k\)q_j
\nonumber\\
&
+
\sum_{j=0}^2\(\beta_j+\sum_{k=0}^2 \alpha_{j+k+1}u_k+\sum_{k=0}^1 \beta_{j+k+1}v_k\)p_j
\nonumber\\
&
-
\sum_{k+\ell\leq 2}(k+1)\mu_{k+\ell+2}(v_k q_\ell-u_k p_\ell),~
\nonumber\\
\tilde{m}(s)
p_0'
&=
\sum_{j=0}^2\(-\gamma_j+\sum_{k=0}^2 \alpha_{j+k+1}w_k+\sum_{k=0}^1 \gamma_{j+k+1}\tilde{v}_k\)q_j
\nonumber\\
&+
\sum_{j=0}^3\(-\alpha_j+\sum_{k=0}^2 \alpha_{j+k+1}\tilde{v}_k+\sum_{k=0}^1 \beta_{j+k+1}w_k\)p_j
\nonumber\\
&-
\sum_{k+\ell\leq 2}(k+1)\mu_{k+\ell+2}(w_k q_\ell-\tilde{v}_k p_\ell),~
\label{longeqsJ}\\
u_j'&=q_0q_j,~~
v_j'=q_0p_j,~~
w_j'=p_0p_j~~(j=0,1,2),
\nonumber\\
q_j&=s^jq_0-\sum_{k+\ell=j-1} (v_k q_\ell-u_k p_\ell)~~(j=1,2,3),
\nonumber\\
p_j&=s^jp_0-\sum_{k+\ell=j-1} (w_k q_\ell-\tilde{v}_k p_\ell)~~(j=1,2,3),
\nonumber\\
\tilde{v}_j&=v_j-\sum_{k+\ell=j-1} (v_k\tilde{v}_\ell-u_k w_\ell)~~(j=1,2),~~\tilde{v}_0=v_0.
\nonumber
\end{align}
Since $\tilde{\bs{K}}_I\to 0$ as $I\to \emptyset$,
boundary values of $q_0, p_0, u_j, v_j, w_j$
at $s\to -1$ are dominated by
the first terms in the Neumann expansion (\ref{Neumann}),

\begin{align}
&q_0(s)\to \tilde{\varphi}(s),~~
p_0(s)\to \tilde{\psi}(s),
\nonumber\\
&u_j(s) \to\int_{-1}^s dx\,x^j \tilde{\varphi}(x)^2,~~
v_j(s) \to \int_{-1}^s dx\,x^j \tilde{\varphi}(x)\tilde{\psi}(x),~~
w_j(s) \to \int_{-1}^s dx\,x^j \tilde{\psi}(x)^2,
\label{bc}
\end{align}
with $\tilde{\varphi}(x)$ and $\tilde{\psi}(x)$ 
given in Eq.\,(\ref{tildephipsi}).
Integrating the diagonal resolvent kernel $R(s)$, 
as determined by Eqs.\,(\ref{longeqsJ}) and (\ref{bc}),
with respect to $s$ gives the Fredholm determinant:
\begin{align}
\det\bigl({1}-\tilde{\bs{K}}_{[-1,s]}\bigr)
=\exp\(-\int_{-1}^s ds'\,R(s')\).
\label{detI-K}
\end{align}

To integrate (\ref{longeqsJ}) to any target value $s\,(>t)$, we use (\ref{bc})
as initial data at $s=-1+\epsilon$ and replace $t$ throughout by $t+i\epsilon'$. 
For suitable positive $\epsilon, \epsilon'$ of order $10^{-10}$, 
the real parts of $q_0(s)$ and $p_0(s)$ are stable and vanish at $s=t$ 
within errors of order $\epsilon$ and $\epsilon'$.
Thus, we can practically circumvent numerical difficulties of (\ref{longeqsJ}) 
at $s=-1$ and $s=t$.

\section{Probability distributions}
\subsection{Joint distribution and ratio distribution}
\begin{figure}[b] 
\centering
\includegraphics[bb=0 0 450 313,width=76mm]{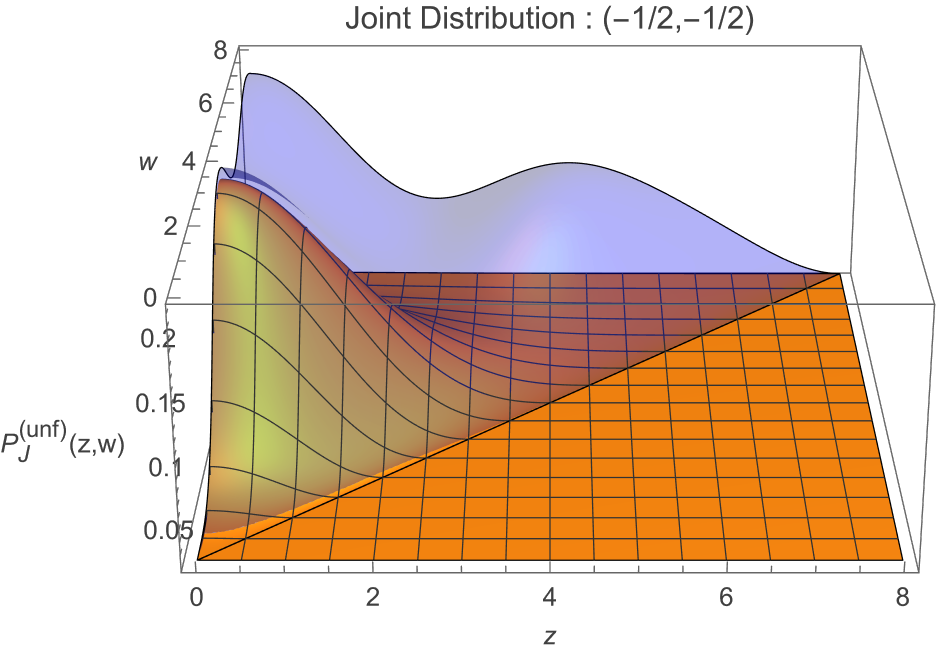}
\includegraphics[bb=0 0 450 313,width=76mm]{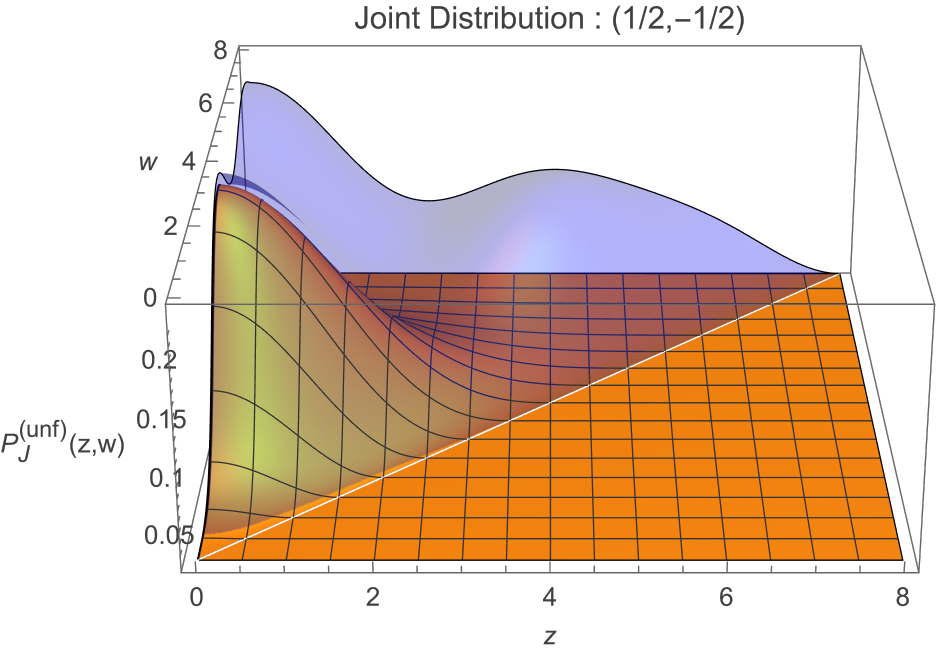}
\includegraphics[bb=0 0 450 313,width=76mm]{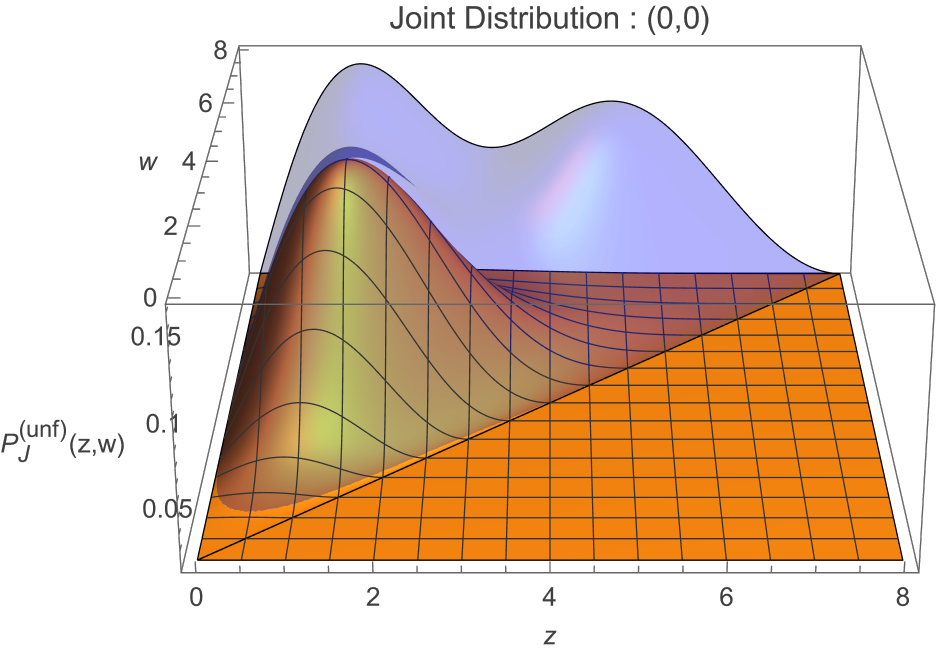}
\includegraphics[bb=0 0 450 313,width=76mm]{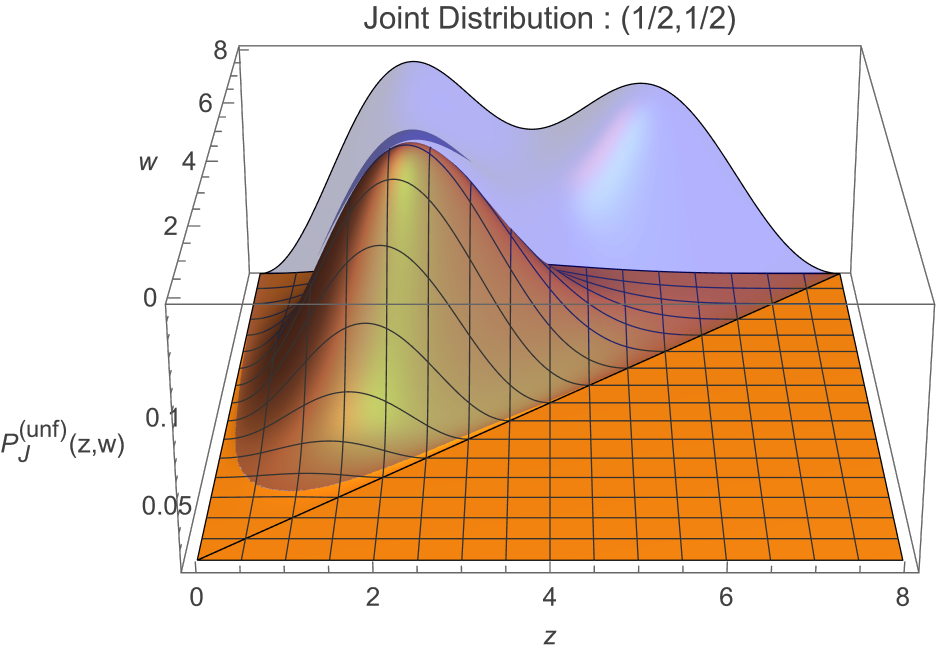}
\caption{
Joint distributions of the two smallest unfolded eigenphases
$P^{(\mathrm{unf})}_{\mathrm{J}}(z, w)$
for the four pairs of indices $(a,b)=(-\frac12, -\frac12), (\frac12, -\frac12), (0,0), (\frac12,\frac12)$
and at $N=12$
shown in orange.
The corresponding unfolded two-point correlation functions
$R_2^{(\mathrm{unf})}(z, w)$ are overlaid in blue.}
\end{figure}
For general $(a,b)$, the TW solution (\ref{mt})--(\ref{detI-K}) gives $P_{\rm J}(t,s)$
via (\ref{rho_Jacobi}) and (\ref{P12}).
Eqs.\,(\ref{Pzw}) and  (\ref{Prr}) then give the unfolded joint distribution
$P_{\rm J}^{(\mathrm{unf})}(z,w)$ (Fig.\,1) and the ratio distribution $p_{\rm J}(r)$ (Fig.\,2).
Their limits are the Bessel joint distribution $P_{\rm B}(z,w)$ \cite{Forrester:1994,Forrester:2007}
(see also Ref.\,\cite{Nishigaki:2021}, Sect. 4.2) 
and its ratio projection $p_{\rm B}(r)$.
As the joint eigenvalue density (\ref{EVD_JUE}) after unfolding contains factors
$\sin(z/N) \bigl(1-\cos(z/N)\bigr)^b$ and
$\bigl(\cos(z/N)-\cos(w/N)\bigr)^2$,
the ratio distribution has the endpoint behaviours $p_{\mathrm{J}}(r) \propto r^{2b+1}$ for $r\to 0$
and $p_{\mathrm{J}}(r)\propto (1-r)^{2}$ for $r\to 1$.
\begin{figure}[tb]
\centering
\includegraphics[bb=0 0 450 308,width=76mm]{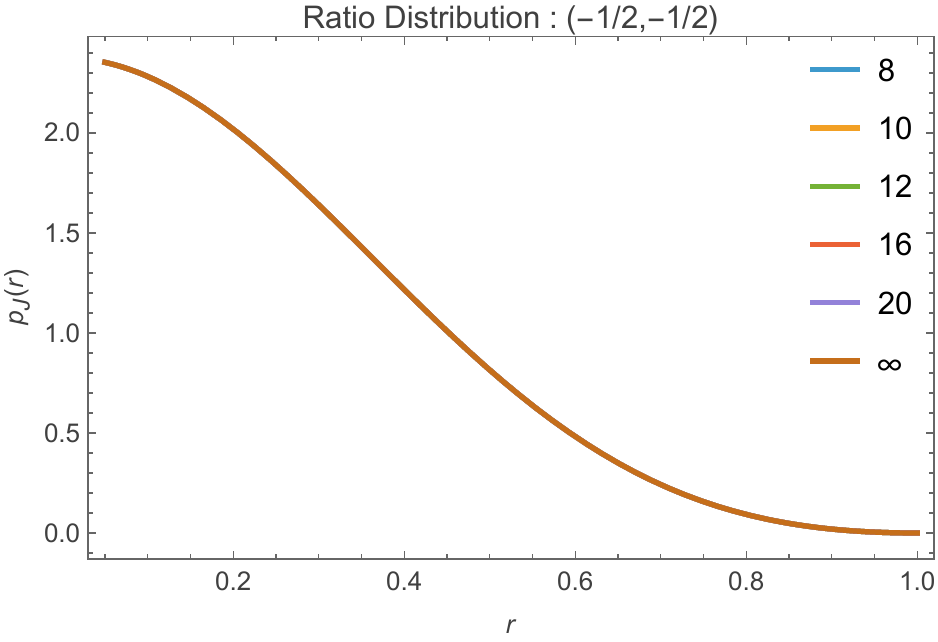}
\includegraphics[bb=0 0 450 314,width=76mm]{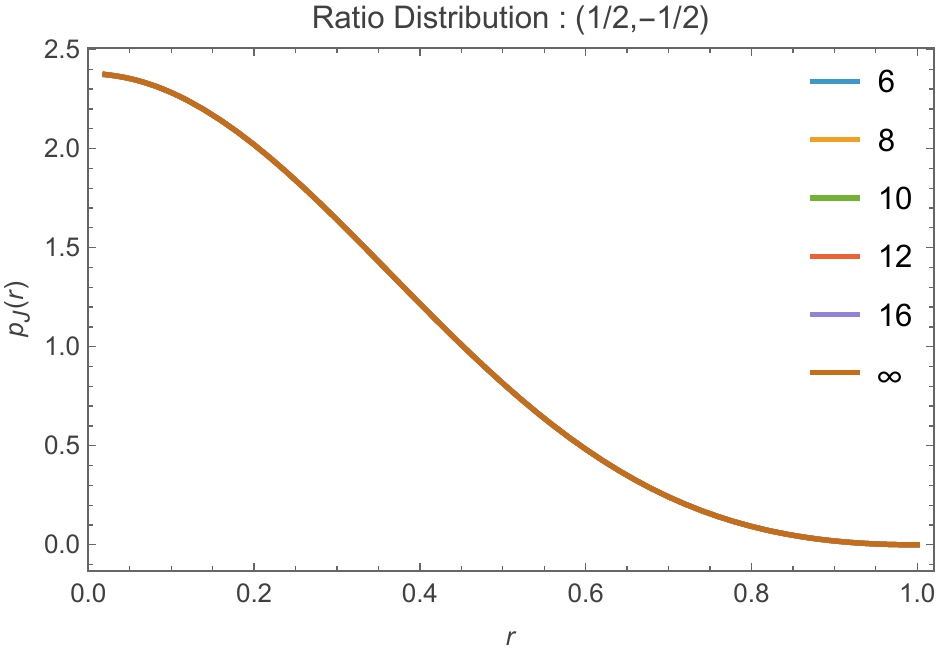}
\includegraphics[bb=0 0 450 308,width=76mm]{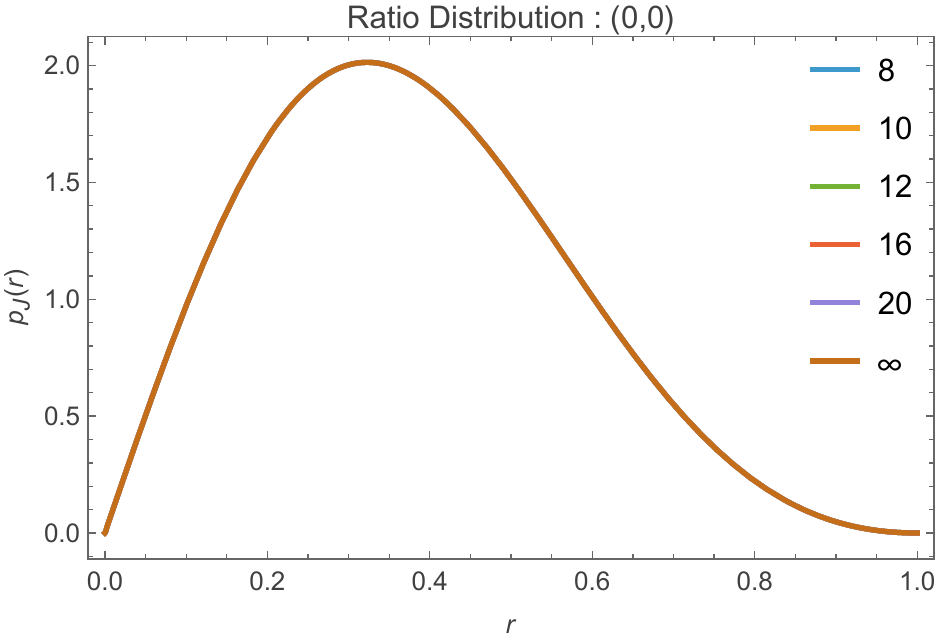}
\includegraphics[bb=0 0 450 308,width=76mm]{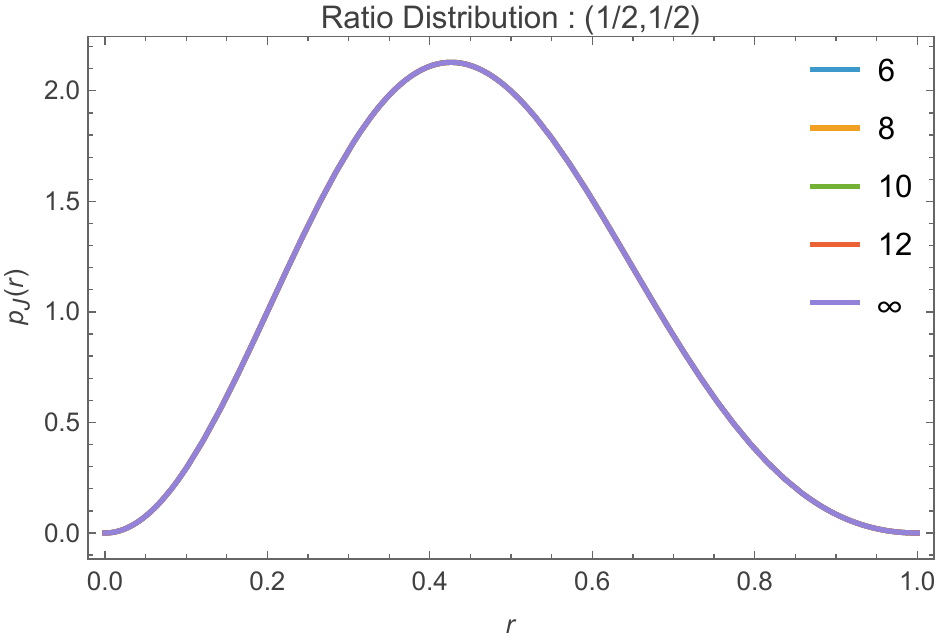}
\caption{
Ratio distributions $p_{\mathrm{J}}(r)$, with $r=z/w$, 
for the two smallest unfolded eigenphases
for the four pairs of indices $(a,b)=(-\frac12, -\frac12), (\frac12, -\frac12), (0,0), (\frac12,\frac12)$,
at several values of $N$ and at $N\to \infty$.
}
\end{figure}
\subsection{Finite-$N$ corrections} 
Even for the smallest values of $N$ (8 or 6), the ratio distribution 
$p_{\mathrm J}(r)$ for JUE$_N$ with any of these index pairs in Fig.\,2
is almost indistinguishable
from its $N\to\infty$ limit $p_{\mathrm B}(r)$
at the corresponding value of $b$.
Remarkably, however, 
the deviations of $p_{\mathrm J}(r)$ from $p_{\mathrm B}(r)$
converge to limiting scaling functions {\em when multiplied by $N^4$} (Fig.\,3).
This empirical finding indicates that, in the expansions 
\begin{align}
P^{\mathrm{(unf)}}_{\mathrm{J}}(z,w)=P_{\mathrm{B}}(z,w)+\sum_{k\geq 1}\frac{1}{N^k}P^{(k)}(z,w),~~
p_{\mathrm{J}}(r)=p_{\mathrm{B}}(r)+\sum_{k\geq 1}\frac{1}{N^k}p^{(k)}(r),
\label{1/Nexp}
\end{align}
the three lower-order corrections
$p^{(1)}(r)$, $p^{(2)}(r)$, and $p^{(3)}(r)$
vanish altogether,
because highly nontrivial cancellations occur when 
$P^{(1)}(z,w)$, $P^{(2)}(z,w)$, and $P^{(3)}(z,w)$
are integrated according to Eq.\,(\ref{Prr}).
This phenomenon parallels that found for bulk spectral statistics in the circular unitary ensemble
(CUE$_N$) \cite{Nishigaki:2026}, 
where the $\mathcal{O}(N^{-2})$ correction to the gap-ratio distribution
relative to the universal sine-kernel limit vanishes, leaving
the $\mathcal{O}(N^{-4})$ term as the leading finite-$N$ correction.%
\footnote{
Using a quadrature approximation \cite{Bornemann:2010a}
for the Fredholm Pfaffian of the reduced Palm kernel
$\tilde{\mathbf{K}}(x,y)=\mathbf{K}(x,y)-\mathbf{K}(x,t)\mathbf{K}(t,t)^{-1}\mathbf{K}(t,y)$,
with $\mathbf{K}(x,y)$ representing the quaternion kernel for the circular
symplectic ensemble (CSE),
we also observed this phenomenon numerically for CSE$_N$.}
The scaled deviations $N^4\bigl(p_{\rm J}(r)-p_{\rm B}(r)\bigr)$
also inherit the endpoint factors 
$r^{2b+1}$ as $r\to0$ and $(1-r)^2$ as $r\to1$.
We also note that the convergence of $N^4 \(p_{\mathrm{J}}(r)-p_{\mathrm{B}}(r)\)$
to $p^{(4)}(r)$ is faster for the cases with $a+b=0$ (upper right and lower left panels in Fig.\,3),
since the next correction, $N^{-1}p^{(5)}(r)$, is absent (see Remark 1).
\begin{figure}[tb]
\centering
\includegraphics[bb=0 0 457 302,width=76mm]{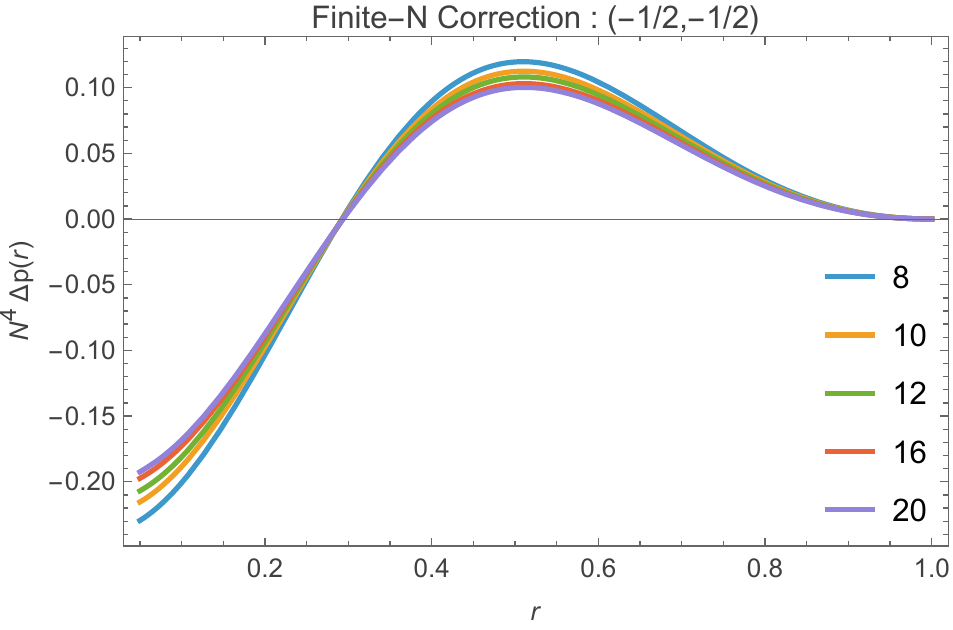}
\includegraphics[bb=0 0 450 293,width=76mm]{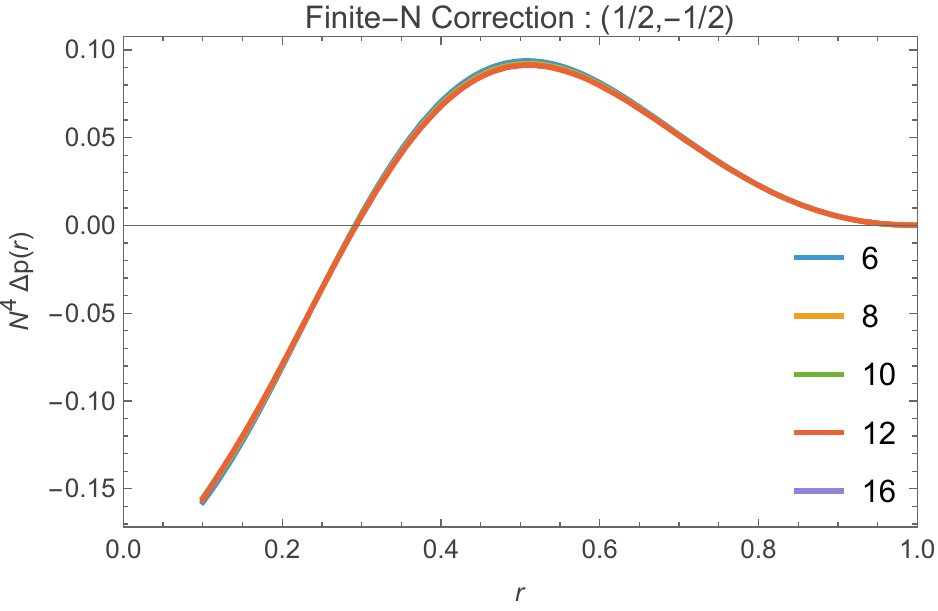}
\includegraphics[bb=0 0 450 308,width=76mm]{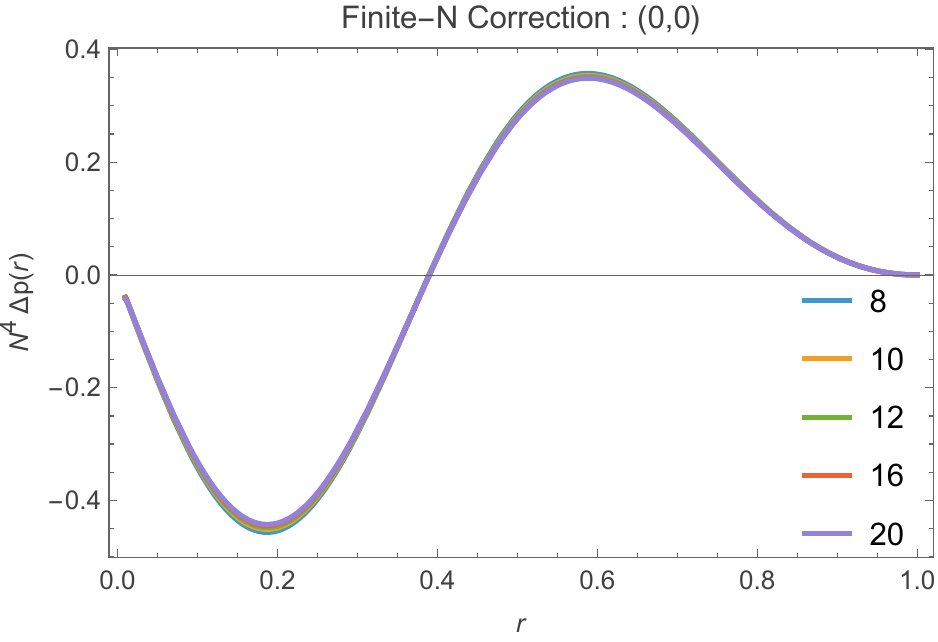}
\includegraphics[bb=0 0 450 302,width=76mm]{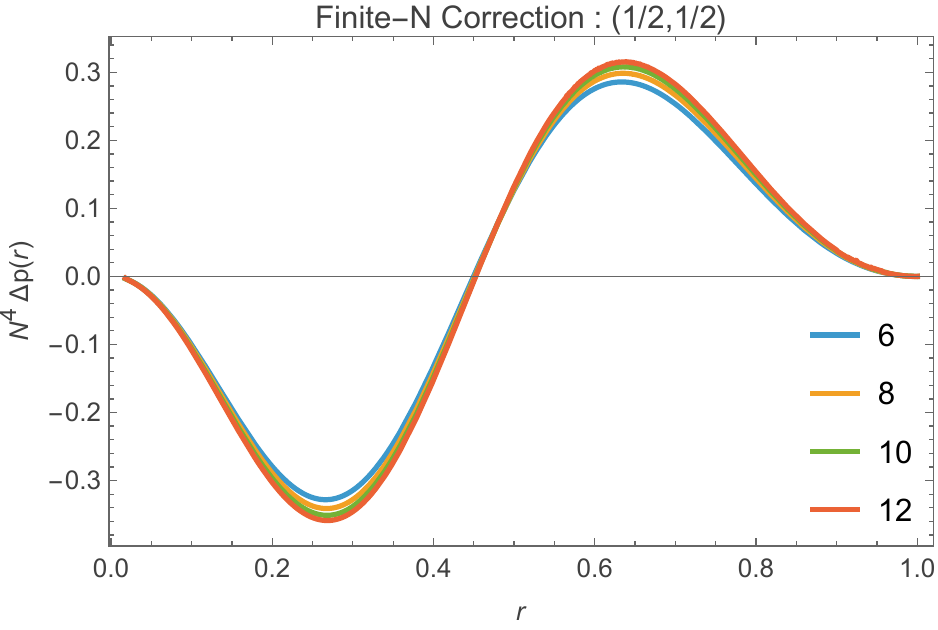}
\caption{
Scaled deviations $N^4\Delta p(r)$ of the ratio distributions 
$p_{\rm J}(r)$ from their $N\to\infty$ limits $p_{\rm B}(r)$,
shown for the four pairs of indices
$(a,b)=(-\frac12, -\frac12), (\frac12, -\frac12), (0,0), (\frac12,\frac12)$
and for several finite values of $N$.
}
\end{figure}

For integer $b$, Eqs.\,(\ref{P12_Jacobi}) and (\ref{Pzw}) allow an analytical check of these cancellations. 
For the simplest Legendre case with $a=b=0$, substitution of 
the improved Mehler-Heine asymptotics  (\ref{impMH}) of the Jacobi polynomials
into Eqs.\,(\ref{P12_Jacobi00}) and (\ref{Pzw}) yields the leading finite-$N$ correction
in the joint distribution,
\begin{align}
&P^{(2)}(z,w)=
\frac{z w}{384} \e^{-{w^2}/{4}}
\times
\nonumber\\
&
 \Bigl\{
    I_3(\sqrt{w^2-z^2}) \left(20 \sqrt{w^2-z^2}
   I_4(\sqrt{w^2-z^2})+(w^4-8 w^2+120)
   I_3(\sqrt{w^2-z^2})\right)
\label{P2zw}\\
&
   -I_2(\sqrt{w^2-z^2}) \Bigl((w^2-8)^2
   I_4(\sqrt{w^2-z^2})+\frac{2 w^4-28 w^2-36 z^2+384
   }{\sqrt{w^2-z^2}}I_3(\sqrt{w^2-z^2})\Bigr)\Bigr\}.
   \nonumber
\end{align}
Upon integrating (\ref{P2zw}) as in (\ref{Prr}), one can confirm,
after some algebra, that
\[
p^{(2)}(r)=
\int_0^\infty
dw\, w\, P^{(2)}(r w, w)=0.
\]
The error caused by replacing the upper limit $N\pi$ with $\infty$ 
is exponentially small in $N^2$ and therefore does not affect 
any algebraic term in expansion (\ref{1/Nexp}).
Thus $p^{(4)}(r)$ is indeed the leading finite-$N$ correction in $p_{\mathrm{J}}(r)$.
For noninteger $b$, however, proving the vanishing of 
$p^{(1)}(r)$, $p^{(2)}(r)$, and $p^{(3)}(r)$ analytically from the TW system 
developed in Sect.\,4 is much more cumbersome; 
we therefore use numerical integration for the representative cases shown in Fig.\,3.
\vspace{3mm}

\noindent
{\it Remark 2.}~ For comparison, consider $N$ independent variables $\{x_k\}$ distributed
according to the Jacobi weight $w(x)$ in (\ref{EVD_JUE}), with $x_k=-\cos(z_k/N)$ as in (\ref{Pzw}). 
An elementary order-statistics calculation gives the limiting ratio distribution $2(b+1)r^{2b+1}$.
The leading finite-$N$ correction is generically of order $N^{-1/(b+1)}$ and proportional to
$r^{2b+1}\left((b+1)/(b+2)-r^2\right),$
with the index $a$ entering only through the amplitude.
This amplitude vanishes when $3a+b+2=0$. In particular, for the Legendre weight
$a=b=0$, the ratio distribution satisfies
\[
p(r)=2r+\frac{4}{3N}(r-2r^3)+O\bigl(\frac{1}{N^{2}}\bigr).
\]
The i.i.d.\,case thus provides a simple comparison with the $N^{-4}$ correction 
found for the JUE.

\section{Dirichlet $L$-function zeros}
The Katz-Sarnak conjecture \cite{Katz:1999} identifies the limiting statistics of
low-lying zeros $\{1/2\pm it_k\}, 0<t_1<t_2<\cdots, $
in families of $L$-functions, averaged over fundamental discriminants $d$,
with those of the low-lying eigenphases of classical compact groups.
In particular, quadratic Dirichlet $L$-functions 
$L(s,\chi_d)=\sum_{n\geq1} \chi_d(n)n^{-s}$, 
where $\chi_d(n)=\left(\frac{d}{n}\right)$ denotes the Kronecker symbol,
correspond to $\mathrm{Sp}(N)$ (class C),
whereas quadratic twists $L(s,\Delta\otimes\chi_d)$ of the Ramanujan $L$-function,
where $\Delta(z)$ denotes the modular discriminant,
correspond to $\mathrm{SO}(2N)$ or $\mathrm{SO}(2N+1)$ (classes D or B, 
according to the root number $\epsilon_d=\pm1$).
Rubinstein \cite{Rubinstein:thesis} provided 
numerical evidence for the symplectic correspondence using quadratic Dirichlet
$L$-functions with $|d|$ up to $10^{12}$.
This hard-edge picture parallels the CUE description of the bulk Riemann zeros associated with 
Montgomery's pair-correlation conjecture \cite{Montgomery:1972,Odlyzko:1987}. 
Finite-height deviations observed by Odlyzko \cite{Odlyzko:2001} agree with finite-$N$ CUE corrections
through order $N^{-2}$, upon setting $N=c\log(t/2\pi)$, but differ at order $N^{-3}$
\cite{Bogomolny:2006,Forrester:2015,Bornemann:2017,Nishigaki:2026}.

For quadratic Dirichlet $L$-functions, restricted-support results
\cite{Ozluk:1993,Ozluk:1999,Fiorilli:2017}
support the symplectic prediction, 
while the ratios conjecture predicts arithmetic lower-order terms \cite{Conrey:2007}.
Since these studies concern the one-level density,
they do not determine the joint distribution of the first two zeros. 
We therefore study the distribution $p_{\mathrm{D}}(r)$ of $r=t_1/t_2$.
As Eq.\,(\ref{Prr}) shows, the ratio distribution can be viewed as a projection
of the joint distribution onto a scale-invariant variable. 
This projection discards information about the absolute scale
while retaining information about the relative positions of the first two zeros.
Arithmetic terms enter the one-level density already at order $N_{\mathrm{eff}}^{-1}$,
where $N_{\mathrm{eff}}(d)=\frac12\log(|d|/2\pi)$, but
whether a correction of this order survives in the ratio distribution is a separate question.

\begin{figure}[tb]
\centering
\includegraphics[bb=0 0 450 308,width=76mm]{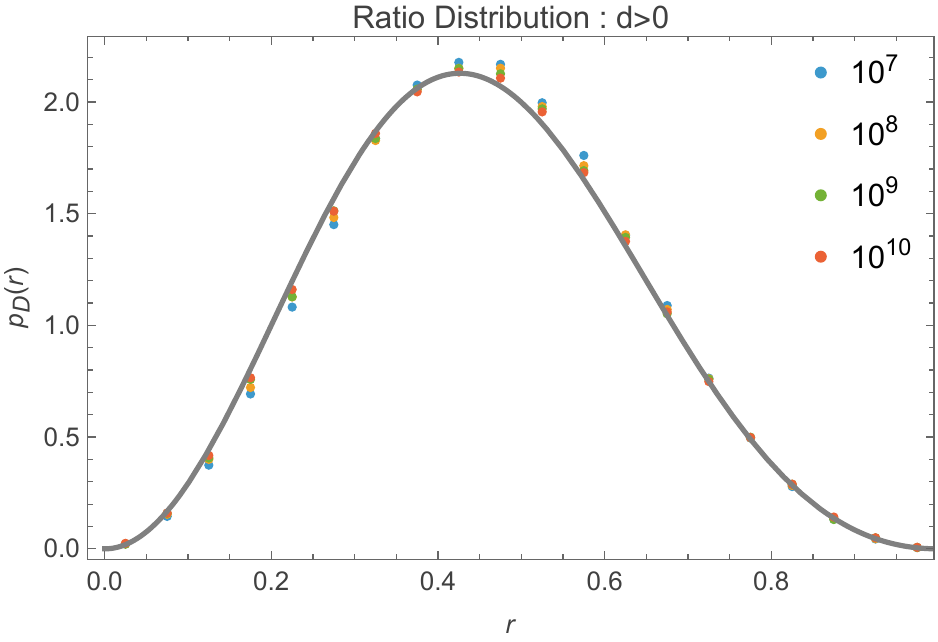}
\includegraphics[bb=0 0 450 308,width=76mm]{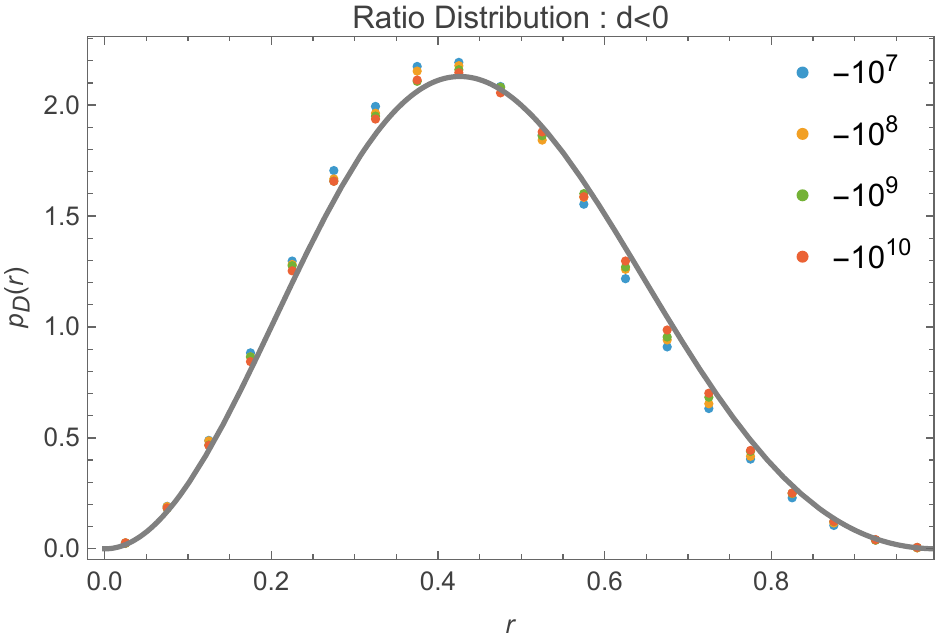}
\includegraphics[bb=0 0 450 307,width=76mm]{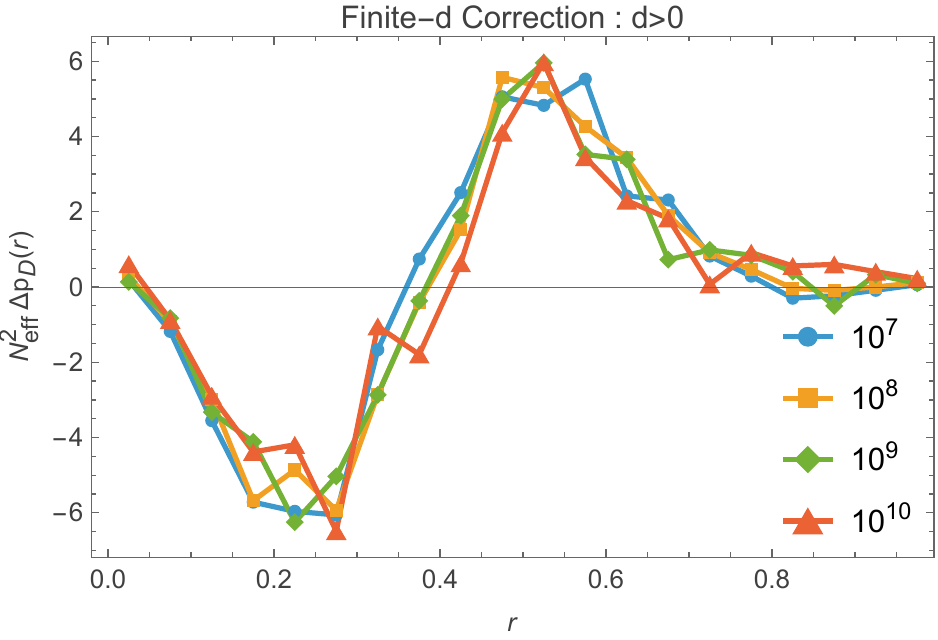}
\includegraphics[bb=0 0 450 307,width=76mm]{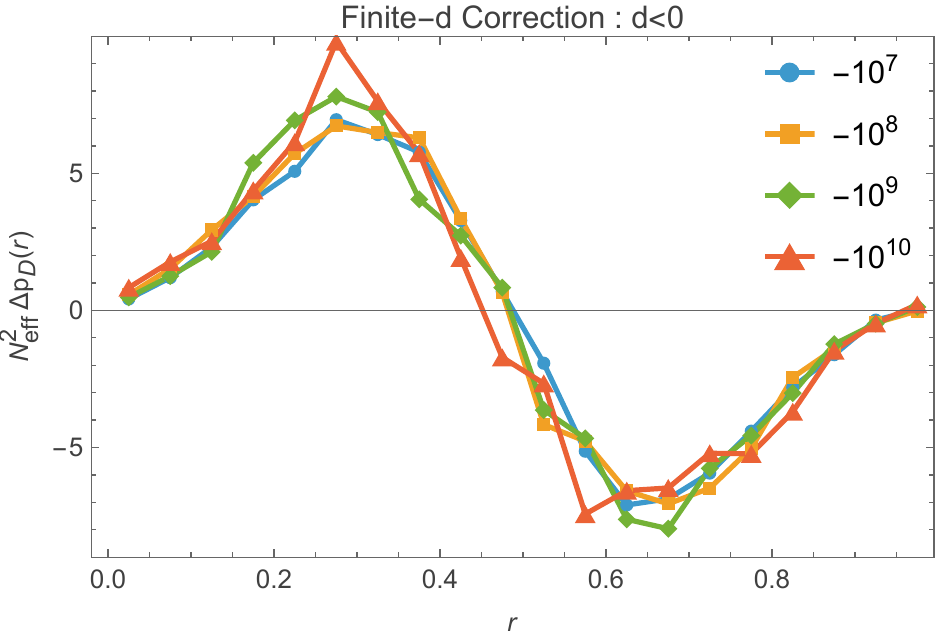}
\caption{
Histograms $p_{\mathrm D}(r)$ of the ratio of the two smallest Dirichlet $L$-function zeros, 
based on approximately 304,000 pairs per window
near $d= 10^7, \ldots,  10^{10}$ (upper left) 
and $d= -10^7, \ldots,  -10^{10}$ (upper right). 
Their scaled deviation from $\mathrm{Sp}(\infty)$ (grey curve), 
$N_{\mathrm{eff}}^2 \Delta p_{\mathrm{D}}(r)$, where
$\Delta p_{\mathrm{D}}(r)=p_{\mathrm{D}}(r)-p_{\mathrm B}(r)$, 
is presented in each lower panel.
}
\end{figure}
To investigate this possibility, 
we computed the first two positive zero ordinates using Stopple's algorithm\footnote{%
The search detects sign changes of the completed $L$-function on the critical line
and does not assume the generalised Riemann hypothesis.}
\cite{Stopple:2009},
separately for positive and negative fundamental discriminants, 
each in four windows
from $10^{7}\leq |d| \leq 10^{7}+10^6$ to $10^{10}\leq |d| \leq 10^{10}+10^6$,
and formed the empirical ratio distributions.
For each window containing approximately 304,000 pairs of zeros,
we compare the resulting distribution $p_{\mathrm{D}}(r)$ with the asymptotic Bessel-kernel result
$p_{\mathrm{B}}(r)$, corresponding to $\mathrm{Sp}(\infty)$ in Fig.\,4, upper panels.
One clearly observes parity-dependent systematic 
deviations from the limiting symplectic density $p_{\mathrm{B}}(r)$.
These deviations scaled by $N_{\mathrm{eff}}(d)^2$,
$N^2_{\mathrm{eff}}\bigl(p_{\mathrm{D}}(r)-p_{\mathrm{B}}(r)\bigr)$,
are plotted in Fig.\,4, lower panels.
They appear to collapse onto two distinct scaling functions for 
positive and negative $d$,
consistent with the absence of an $N_{\rm eff}^{-1}$ term in the expansion
\begin{align}
p_{\mathrm{D}}(r)=p_{\mathrm{B}}(r)+
\sum_{k\geq 1}\frac{1}{N_{\mathrm{eff}}^k} p_{\mathrm{D}}^{(k)}(r).
\label{1/Neffexp}
\end{align}
Integrating the limiting J\'{a}nossy density evaluated by Nystr\"{o}m quadrature
\cite{Bornemann:2010a} gives $\mathbb{E}[r]_{\mathrm{Sp}}=0.450249256332009797(1)$. 
The deviations of the empirical means from this value, 
plotted against $N_{\mathrm{eff}}(d)^{-2}$ in Fig.\,5, are also consistent with this scaling.
\begin{figure}[tb]
\centering
\includegraphics[bb=0 0 450 276,width=75mm]{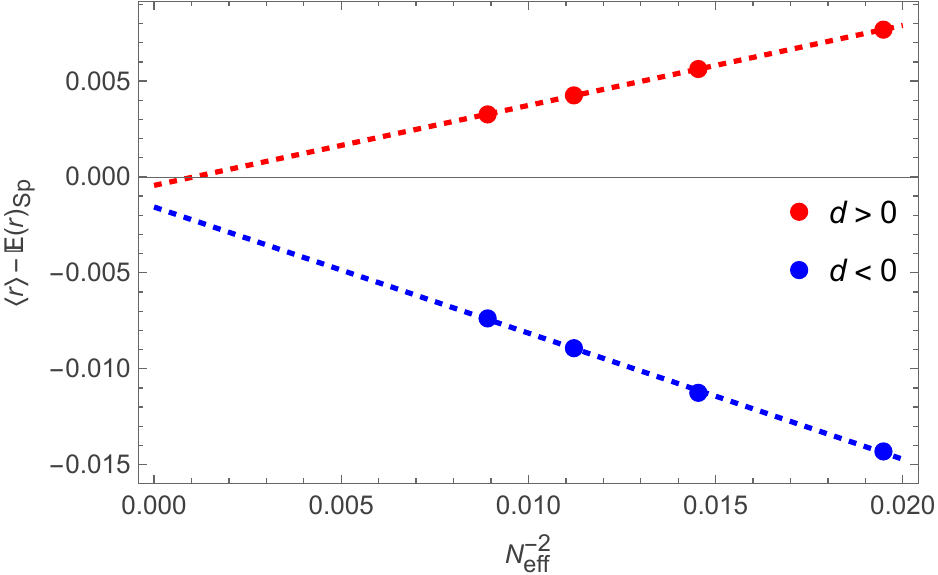}
\caption{
Deviations of 
the mean ratio $\langle r\rangle$ of the two smallest Dirichlet $L$-function zeros
in each window of $d$
from the $\mathrm{Sp}(\infty)$ prediction $\mathbb{E}[r]_{\mathrm{Sp}}$, 
as functions of $N^{-2}_{\mathrm{eff}}$.
The dotted lines are linear fits with free intercepts
to each set of four points, shown as guides to the eye.}
\end{figure}
These numerical results point to an unexpected agreement with $\mathrm{Sp}(N)$: 
the first-order coefficients in (\ref{1/Nexp}) and (\ref{1/Neffexp}) both appear to vanish. 
With $N$ identified with $N_{\rm eff}$, 
the ratio distributions differ at order $N_{\rm eff}^{-2}$,
i.e.\, $p_{\mathrm{D}}^{(2)}(r)$ is nonvanishing and parity-dependent,
whereas $p^{(2)}(r)$ vanishes.

\section{Concluding remarks}
Our finding that, for the cases considered, the leading finite-$N$ correction
 in the distribution of the ratio
of the two smallest unfolded eigenphases of JUE$_N$ is of order $N^{-4}$
suggests that, in ensembles of unitary matrices 
arising in quantum-chaotic models with chiral or particle-hole symmetry,
eigenphase ratios may reveal subtle spectral information that would otherwise be inaccessible.
Potential applications include chiral-symmetric Floquet operators in quantum-chaotic systems%
\footnote{
A concrete example of disordered class-AIII Floquet settings is 
the SSH-type loop drive \cite{Culver:2025},
although whether its eigenphase statistics follow the JUE remains to be investigated.}
and symmetry-constrained unitary circuits on bipartite lattices.
Consider an ensemble of such unitary matrices with $N$ pairs of eigenphases $\{\pm\theta_j\}$,
whose local correlations near $\theta=0$ or $\pi$ converge, as $N\to\infty$,
to those of the limiting Bessel kernel.
If their finite-size correction to the one-point density or the individual distributions agrees
with that of JUE$_N$, the universality of random matrix theory is extended to 
those quantum-chaotic models to order $N^{-1}$.
Agreement of the finite-size correction to the ratio distribution would
serve as an indication of the extended universality through order $N^{-4}$, 
including the absence of lower-order terms.

\section*{Data availability}
The computed ordinates of the first two positive zeros of the quadratic Dirichlet $L$-functions
studied here are available on Zenodo {\tt 10.5281/zenodo.24536334}.
The numerical grids for the joint eigenphase densities $P^{(\mathrm{unf})}_{\mathrm{J}}(z,w)$
are available from the author upon request.

\section*{Acknowledgments}
This work was first presented at the conference 
`{\bf G}athering {\bf A}dvances in Random Matrices and Beyond'
 (Institut Henri Poincar\'{e}, Paris, June 2026). 
 I thank the organisers for the invitation and the participants for stimulating discussions.

\appendix
\renewcommand{\thesection}{Appendix}
\section{Jacobi polynomials}
The Jacobi polynomials $\{P_n^{(a,b)}(x)\}_{n=0,1,2,\ldots}$
are defined to be orthogonal with respect to the weight $w(x)$ in
(\ref{EVD_JUE}),
with leading coefficients and norms given by
\begin{align}
&P_n^{(a,b)}(x)=c_{n}^{(a,b)} x^n+\cdots,~~~
\int_{-1}^1 dx\,
(1-x)^a (1+x)^b
P_n^{(a,b)}(x)P_m^{(a,b)}(x)=h^{(a,b)}_n\delta_{nm},
\nonumber\\
&c_{n}^{(a,b)}=\frac{\Gamma(2n+a+b+1)}{2^n n! \,\Gamma(n+a+b+1)},
~~~
h^{(a,b)}_n=\frac{2^{a+b+1} \Gamma (n+a+1) \Gamma (n+b+1)}{n! (2n+a+b+1) \Gamma (n+a+b+1)},
\label{norm}
\end{align}
with removable singularities interpreted by continuity.
They  satisfy a second-order linear differential equation,
\begin{align}
\left\{(1-x^2)\frac{d^2}{dx^2}-\(a-b+(a+b+2)x\)\frac{d}{dx}+n(n+a+b+1)
\right\} P_n^{(a,b)}(x)=0.
\label{SLeq}
\end{align}
We introduce $N=n+(a+b+1)/2$ and set $x=-\cos(z/N)$. 
In these variables, we expand Eq.\,(\ref{SLeq}) in powers of $N^{-2}$ at fixed $z$.
Recursively solving each order of this expansion in terms of the Bessel functions, one obtains
the improved Mehler-Heine asymptotics of the Jacobi polynomials
to any desired order  \cite{Frenzen:1985}. Here we present only the leading $1/N^2$ correction:
\begin{align}
&\frac{(-1)^n n!}{\Gamma (n+b+1)}
P^{(a,b)}_n\left(-\cos\frac{z}{N}\right)=
\nonumber\\
&
\left(\frac{z}{2}\right)^{-b} \left\{
J_b(z)+\frac{  (3 a+b+2) z^2 J_b(z)+ (3 a^2+b^2-1) z\,  J_{b+1}(z)}{24N^2}
+O\Bigl(\frac1{N^4}\Bigr)
\right\}.
\label{impMH}
\end{align}


\end{document}